\documentclass[aps,preprint,superscriptaddress]{revtex4}
\usepackage{graphicx} 
\usepackage{mathtools}
\usepackage{mathptmx}

\begin{document}
\title{Structural transition of vortices to nonlinear regimes in a dusty plasma}
\author{Modhuchandra Laishram}
\address{CAS Key Laboratory of Geospace Environment and Department of Engineering 
and Applied Physics, University of Science and Technology of China, Hefei 230026,
China}
%
\author{Ping Zhu}
\address{CAS Key Laboratory of Geospace Environment and Department of Engineering 
and Applied Physics, University of Science and Technology of China, Hefei 230026,
China}
\address{KTX Laboratory and Department of Engineering and Applied Physics, 
University of Science and Technology of China, Hefei 230026, China}
\address{Department of Engineering Physics, University of Wisconsin-Madison, 
Madison, Wisconsin 53706, USA}
\date{\today}

\begin{abstract}
A 2D hydrodynamical model is developed and analyzed for the steady 
state of a driven-dissipative dust clouds confined in an azimuthally symmetric 
toroidal system which is in dynamic equilibrium with background unbounded 
streaming plasma. Its numerical solution not only confirms the analytical structure 
of the driven dust vortex flow in linear limit as reported in previous analysis, but 
also shows how the dust vortices are strongly affected by the nonlinear 
convection of the flow itself. Effects of various system 
parameters including external driving field and Reynolds number (Re) are investigated 
within the linear to nonlinear transition regime $0.001\le {\rm Re} < 50$. 
In agreement with various relevant experimental 
observations, the flow structure which is symmetric around center in the linear regime 
begins to turn asymmetric in the nonlinear regime. The equilibrium structure of dust 
flow is found to be influenced mainly by the dissipation scales due to 
kinematic viscosity, ion drag, and neutral collision in the nonlinear regime, 
whereas in the linear regime, it is mainly controlled by the external 
driving field and the confining boundaries. 
\end{abstract}



\maketitle

\section{Introduction}
\label{introduction}
Vortex or swirling flows about an axis are ubiquitous in most complex
fluids involving multiple spacial and temporal scales. The key feature 
of such complex fluids is the interactions of various constituents that lead to 
collective behaviors such as self-organization and pattern 
formation~\cite{PhysRevE.81.041124,RevModPhys.72.603,PhysRevE.91.063110,
doi:10.1063/1.4887003}. For examples, flows in polymer solutions, colloidal 
suspensions and many biological systems are complex flows in low Reynolds number
regimes that exhibit vortex or circulating flow 
pattern~\cite{lauga,Wolgemuth20081564,PhysRevLett.110.174502,
Yager,RevModPhys.77.977,Stroock647}.
On the other hand, similar self-organized vortex flows are also observed
in many higher Reynolds number macroscopic flow systems such as 
White Ovals, Great Red Spot of Jupiter, polar vortex of Earth, and 
macroscopic circulations including hurricanes and tornadoes in planetary 
atmospheres~\cite{Sommeria,CHOI200735,marcus_1990}. The vortex structure in such driven 
dissipative systems evolves in various dynamical regimes. It has been observed in different 
sizes, shapes, orientations, aspect ratios, and convection velocities. 
It has been recognized that vortex structure plays an important role in fluid mixing and 
transport process in laminar and turbulent flows~\cite{Nivedita,marcus_1990}. 
Thus it has been a topic of active research that how the vortical structure evolution in 
various complex fluid flows depends on system parameters, in particular the 
Reynolds number from the viscous dominant linear regime to inertial 
dominant nonlinear flow regime. When the flow is viscous dominant which happen mostly 
at lower flow velocity and smaller length scale (Re$ \le 1$),
the gradients in the hydrostatic pressure are balanced by viscous diffusive
producing smooth deformation and laminar flows. However, at higher flow velocity 
and larger length scale (Re$\gg 1$), the viscous stress is comparable or 
less than the inertial convective transport and the pressure gradients act to 
accelerate the fluid elements and generate nonlinear convective flows.

Interestingly, dust clouds electrostatically suspended in a plasma can 
be a realizable prototype for experimental or theoretical formulation for various 
characteristics of driven-dissipative complex fluid flows. 
Dusty plasmas with coupling parameters in the ranges of $10\le \Gamma \le 100$ for a
fixed screening parameter $\kappa$ behave like complex fluids. They can form 
many self-organized vortices or circulating flow as collective behaviors due 
to its long-range Coulomb interactions~\cite{Fortov2003,PhysRevLett.83.1598}. 
Here $\Gamma = E_{p}/E_{k}$, i.e., the strength of Coulomb potential energy $E_{p}$ over 
kinetic energy $E_{k}$ and the screening parameter $\kappa = d/\lambda_D$, i.e., the ratio 
of the particle distance over the Debye length due to background plasma
~\cite{PhysRevLett.73.652,PhysRevLett.83.1598}. 
Study of dust vortex structures in such a complex plasma  
presents an attractive option for analyzing and interpreting the dynamics of 
many relevant complex fluid flows in laboratories, industrial applications as well 
as many natural processes. In the previous linear (Re$ \ll 1$) 
analysis~\cite{PhysRevE.91.063110,doi:10.1063/1.4887003}, 
the analytic structure of the steady dust flow driven by an unbounded 
and sheared ion flow is obtained.  
The scales of the vortex developed in linear regime mainly 
depend on those of the ion flow driver, the boundaries, and as 
various system parameters including the kinematic viscosity. 
In recent dusty plasma 
experiments~\cite{doi:10.1063/1.4929916,doi:10.1063/1.4916065}, 
a very localized and isolated regions of acceleration and frictional 
retardation in velocity field, along with an uniform vorticity region surrounded by 
highly sheared layers of dust vortex motion are produced. This suggests that 
the laboratory dusty plasma may have entered the nonlinear regime of the vortex flow 
and lends itself to the study of macroscopic nonlinear dynamics in such systems of 
similar Reynolds number~\cite{PhysRevE.91.063110}. 
It motivates us to extend
the previous linear analysis to higher Reynolds number flow regime (Re$\gg 1$),
where the nonlinear inertial flow is effective which may explain many new flow 
characteristics as observed in dusty plasma experiments~\cite{doi:10.1063/1.4929916}.

In present work, we employ a 2D numerical model 
for the dynamics of a steady isothermal dust fluid confined in an axisymmetric toroidal 
configuration in dynamical equilibrium with an unbounded sheared flow of a streaming ions 
through a combination of electrostatic and gravitational fields. 
This model is very general and is applicable to any driven dissipative systems in arbitrary 
Reynolds number flow regimes.  
The dust fluid is assumed 
to be incompressible, viscous, and Newtonian, in which the dust is dragged by the shear 
streaming of ions through the confined domain. In such a system where an external 
momentum source is present, specific steady flow solutions 
are attainable only in presence of a frictional sink of momentum which is afforded by the 
stationary background plasma and neutral fluids. The main objective of present work is 
to understand the characteristics of dust flow in presence of inertial 
flow at higher Reynolds number regimes and the corresponding changes in vortex 
structure from linear to nonlinear flow regimes. 

The manuscript is organized as follows. The 2D hydrodynamic 
model and numerical methods for studying the dynamics of 
driven-dissipative dust fluid for arbitrary Re in a toroidal configuration are 
introduced in Sec.~\ref{formulation}. Then a detailed characterizations of the 
steady dust flow structures in linear and nonlinear regime are discussed in 
Sec.~\ref{characterization}.
The nonlinear effects on the vortex structure with varying 
system parameters, namely, the external driving field, the kinetic viscosity of the 
dust fluid $\mu$, the ion dragging co-efficient $\xi$, and the neutral collision 
frequency $\nu$ are described in Sec.~\ref{analysis_2} and Sec.~\ref{analysis_3} 
respectively. Summary and conclusions are presented in Sec.~\ref{conclusion}.

\section{Hydrodynamic model and numerical method}
\label{formulation}
\begin{figure}
\includegraphics{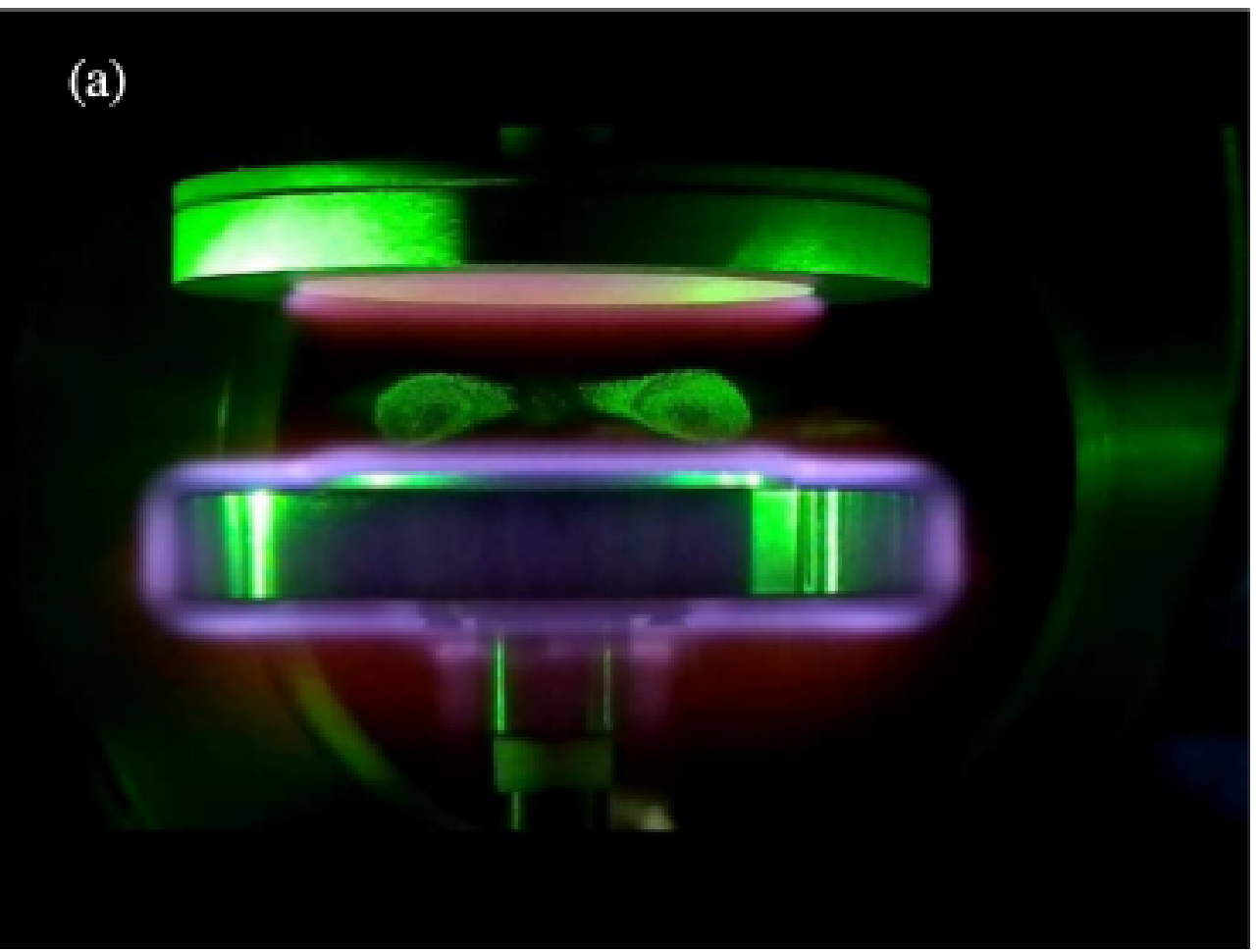}
\includegraphics{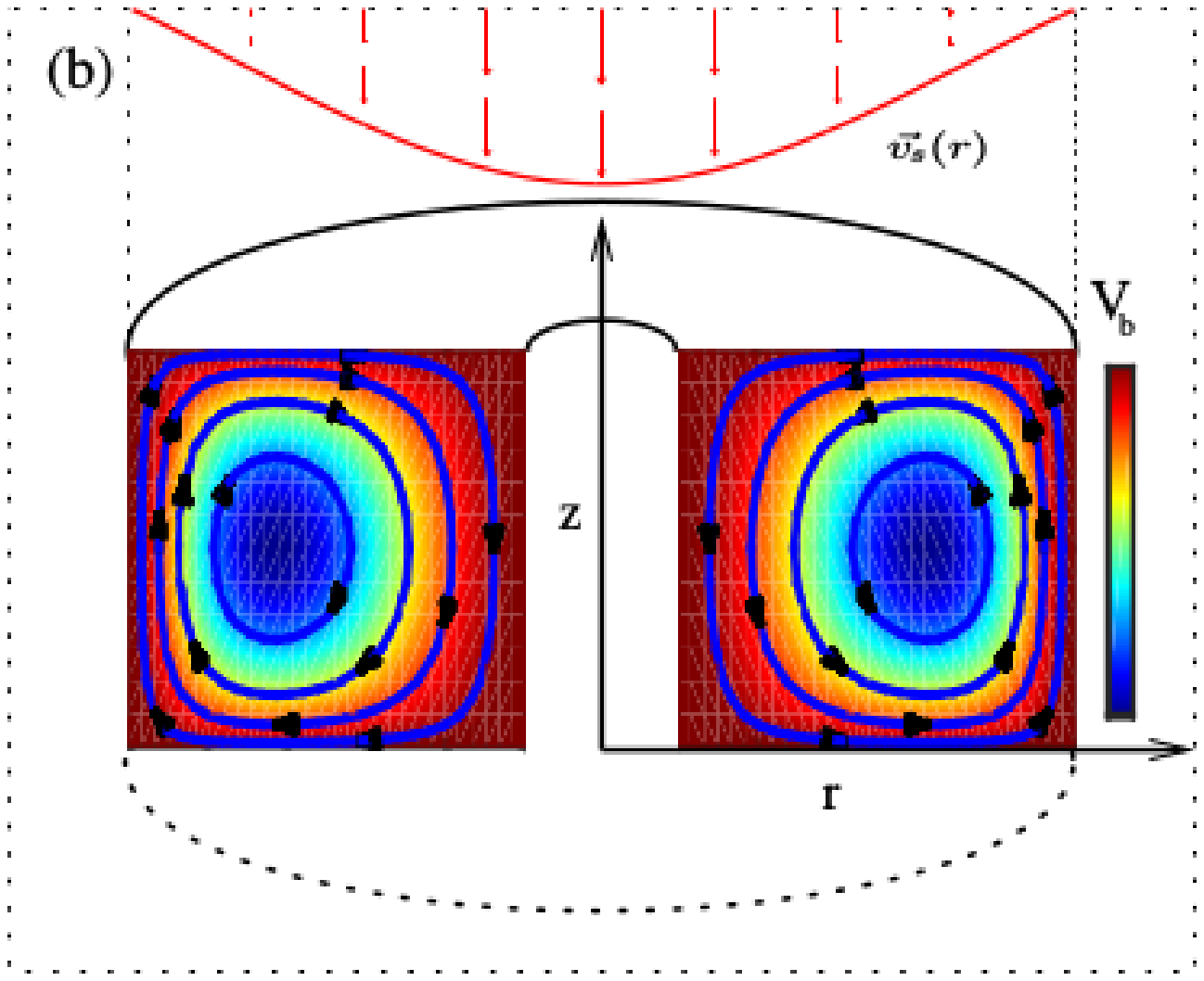}
\caption{(a) Toroidal dust clouds with poloidal circulation 
in the laboratory dusty plasma experimental setup by 
M.~Kaur{\it ~et~ al.}~\cite{doi:10.1063/1.4929916}.
(b) Schematic representation of dust cloud confined toroidally 
by the electrostatic potential ${V_b(r,z)}$ and driven poloidally 
by the sheared ion flow {$\upsilon_s(r,z)$} through out the dust clouds 
[Reprinted courtesy of AIP Conference Proceedings {\bf 1925}, 020028 (2018) 
by permission]
 \label{schematic}}
\end{figure}

The geometry of confined dust fluid is taken from the recent experiment 
where a toroidal dust cloud flowing in poloidal direction in a glow discharge 
plasma, as shown in Fig.~\ref{schematic}(a)~\cite{doi:10.1063/1.4916065,
doi:10.1063/1.4929916}. 
The confinement and localization of the dust clouds in these and similar 
laboratory experiments are achieved effectively by a combination of 
electrostatic and gravitational fields, where a 2D or 3D conservative
field ${\bf F_{c}}=-\nabla V_b$ is prepared with various experimental 
means 
\cite{PhysRevB.61.8404,PhysRevLett.101.125002,0295-5075-102-4-45001,
doi:10.1063/1.2147000}.
The dust fills the volume of the torus where the shape of its poloidal cross-section is 
determined by the effective confining potential. 
However, for the computational simplicity, the system in present model is 
approximated as being axisymmetric and the poloidal cross-section of the torus is 
simplified as a rectangle as illustrated schematically in Fig~\ref{schematic}(b). 
The toroidal dust fluid is considered confined by an effective potential $V_b(r,z)$ within 
the boundaries of a finite section of an infinitely long cylinder of flowing plasma, 
where $0<r/L_{r}<1$, $-1<z/L_{z}<1$, and $0<\phi<2\pi$. 
The effective confining potential ${V_b(r,z)}$ jumps from a small 
value within the domain to a very high value at the rectangular boundary, which serves as 
a rectangular equipotential line for the boundary condition of a perfect confinement.

For the dust flow that satisfies 
incompressible 
and isothermal conditions 
and has a finite viscosity, the dynamics is governed by the simplified
Navier-Stokes equations in which the drive produced by the sheared ion 
drag and the friction produced by the stationary background neutral fluid 
can be accounted for non-conservative vorticity sources~\cite{landau2013fluid},
\begin{eqnarray}
\nonumber
\frac{\partial {\bf u}}{\partial t}  + ({\bf u} \cdot \nabla) {\bf u}  
= - \frac{\nabla{P}}{\rho} - \nabla{V}  + \mu \nabla^2 {\bf u}    
\hspace{1.4cm} \\  
 - \xi ({\bf u} - {\bf v})  - \nu ({\bf u} - {\bf w}).
\label{ns-equation}
\end{eqnarray}
Here ${\bf u}$, ${\bf v}$ and ${\bf w}$ are the flow 
velocities of the dust, ion, and neutral fluids, respectively. 
$P$ and $\rho$ are the pressure 
and mass density of the dust fluid, respectively, $V(r,z)$ is the confining 
potential, $\mu$ is the kinematic viscosity, $\xi$ is the coefficient of ion drag 
acting on the dust, and $\nu$ is the coefficient of 
friction generated by the stationary neutral fluid 
\cite{PhysRevLett.68.313,PhysRevE.66.046414,PhysRevLett.92.205007}.
The overall combination of charged dust and background plasma is 
quasi-neutral and the electrons are in thermal equilibrium with the 
streaming ions and the confined dust. The incompressibility condition of the
confined dust fluid 
\begin{eqnarray}
\nabla\cdot {\bf u}=0,
\label{continuity}
\end{eqnarray}
defines a potential called stream function ${\Psi}$ such that 
${\bf u}=\nabla\times{\Psi}$.
In case of an azimuthally symmetric dust flow the dust motion can
suitably be treated in the 2-dimensional $r$-$z$ plane where the
dust vorticity vector $\vec{\omega}=\nabla\times{\bf u}$ is in the 
toroidal direction only. Then the stream function becomes like a scalar 
field for two dimensional flows giving 
${\bf u}=\nabla\times(\psi\hat{\phi})$. 
 For a stationary background neutral fluid (${\bf w}=0$), the equations for the 
two-dimensional steady dust flow can be derived from 
Eqs.~(\ref{ns-equation})-(\ref{continuity}) as follows,
\begin{eqnarray} 
\nabla^2\psi&=&-\omega,
\label{streamfunction-equation}\\
({\bf u} \cdot \nabla) {\omega}&=&\mu \nabla^2 \omega - 
(\xi+\nu) \omega +\xi \omega_s,  
\label{vorticity-equation}
\end{eqnarray}
where $\omega_{s}$ is the external vorticity source from the unbounded
sheared ion flow. 
The 2D solutions of 
(\ref{streamfunction-equation})-(\ref{vorticity-equation}) were recently
obtained in the linear regime (Re$<1$), 
where the inertial effects are dominated by the diffusive transport 
and thus the nonlinear term in the left-hand side of the momentum equation
can be ignored in the linear viscous 
limit~\cite{doi:10.1063/1.4887003,PhysRevE.91.063110}. 
The Eqs.~(\ref{vorticity-equation}) in the linear limit
admits standard solution procedures where integration is possible for 
an individual mode of the dust vorticity interacting with that of the 
driver. As presented in Ref.~\cite{doi:10.1063/1.4887003,PhysRevE.91.063110}, 
such 2D linear solutions are obtained by constructing an eigenvalue problem and 
representing the dust and source stream function in terms of a set of 
orthogonal eigenfunctions that satisfy the appropriate boundary conditions.
In nonlinear regime, the equations can be solved using proper numerical approach. 
Here the method of Successive Over Relaxation (SOR)~\cite{press1992art} is adopted 
in the present study.

In the first step, the Eqs.~(\ref{streamfunction-equation}) and
Eqs.~(\ref{vorticity-equation}) are normalized 
using proper scaling units $L_r$ and $U_0$ etc, such that the 
$\psi\rightarrow\psi/[U_0L_r]$, $\omega\rightarrow\omega/[U_0/L_r]$,
${\bf u}\rightarrow{\bf u}/[U_0]$, $\omega_s\rightarrow\omega_s/[U_0/L_r]$, 
$\mu\rightarrow\mu/[U_0L_r]$, $\xi\rightarrow\xi/[U_0/L_r]$, 
and $\nu\rightarrow\nu/[U_0/L_r]$, etc. 
Then the relative importance of various
terms in the equation can be compared in terms of the magnitude of the 
dimensionless coefficients. 
Now, in normalized dimensionless form, the 2-D steady incompressible N-S equations 
for the dust fluid can be written as
\begin{eqnarray}
0&=& \nabla^2\psi  + \omega,
\label{stream-equation1}\\
0&=& \nabla^2 \omega 
-\frac{1}{\mu}({\bf u} \cdot \nabla) {\omega}  
-\frac{ (\xi+\nu)}{\mu} \omega 
+\frac{\xi}{\mu} \omega_s.
\label{vorticity-equation1}
\end{eqnarray}
Then the equations are cast in the form suitable for
numerical solutions. Using cylindrical coordinates $(r,\phi,z)$ and assuming axisymmetry, 
Eqs. (\ref{stream-equation1}) and (\ref{vorticity-equation1}) can be solved in iterative 
steps,
\begin{eqnarray}
\nonumber
\psi^{n+1}=\psi^{n}+\Delta L^2\left(\frac{\partial^2 }{\partial r^2}
+\frac{1}{r}\frac{\partial }{\partial r} -\frac{1}{r^2}
+ \frac{\partial^2 }{\partial z^2} \right)\psi^{n} \\ 
+ {\Delta L^{2}}\omega^{n},
\label{psi-equation}\\  \nonumber
 \omega^{n+1}=\omega^{n} +\Delta L^2\left(\frac{\partial^2 }{\partial r^2}
+\frac{1}{r}\frac{\partial }{\partial r} -\frac{1}{r^2}		
+ \frac{\partial^2}{\partial z^2} \right)\omega^{n} \\ \nonumber
- \frac{\Delta L^{2}}{\mu}\left(u_r\frac{\partial \omega}{\partial r}\right)^{n}
- \frac{\Delta L^{2}}{\mu}\left(u_z\frac{\partial \omega}{\partial z}\right)^{n}\\ 
- {\Delta L^{2}}K_1 \omega^{n}
+ {\Delta L^{2}}K_2 \omega_s,
\label{omega-equation}
\end{eqnarray}
where $\Delta \psi \equiv \psi^{n+1}-\psi^{n}$ is iterative variation, ${\Delta L}$ is the 
iteration step size (or virtual time step size), $K_{1}=(\xi+\nu)/\mu$, $K_{2}=\xi/\mu$, 
and $n$ represents the iteration step. Since $\psi^{n+1}\rightarrow \psi^{n}$ for 
the steady state, the above equations can be rearranged as follows,
\begin{eqnarray}
\nonumber
\left[1-
\Delta L^2\left(\frac{\partial^2 }{\partial r^2}
+\frac{1}{r}\frac{\partial }{\partial r} 
+ \frac{\partial^2}{\partial z^2} \right)\right]
\psi^{n+1}=\psi^{n}\\ 
+ {\Delta L^{2}}\omega^{n}-\frac{\Delta L^{2}}{r^2}\psi^{n},
\label{psi-equation1}\\\nonumber
\left[1-
\Delta L^2\left(\frac{\partial^2 }{\partial r^2}
+\frac{1}{r}\frac{\partial }{\partial r} 
+ \frac{\partial^2}{\partial z^2} \right)\right] 
\omega^{n+1}=\omega^{n} 
\\ \nonumber
- \frac{\Delta L^{2}}{\mu}\left(u_r\frac{\partial \omega}{\partial r}\right)^{n}
- \frac{\Delta L^{2}}{\mu}\left(u_z\frac{\partial \omega}{\partial z}\right)^{n}
\\ 
-\frac{\Delta L^{2}}{r^2}\omega^{n}
- {\Delta L^{2}}K_1 \omega^{n}
+ {\Delta L^{2}}K_2 \omega_s.
\label{omega-equation1}
\end{eqnarray}
For numerical efficiency, the above equations
are rewritten in a two-operator form where each operator has only one directional 
derivative as follows,
{\small
\begin{eqnarray}
\nonumber
\left[1-
\Delta L^2\left(\frac{\partial^2 }{\partial r^2}
+\frac{1}{r}\frac{\partial }{\partial r} 
\right)
\right]
\left[1-
\Delta L^2\frac{\partial^2}{\partial z^2} 
\right]
\psi^{n+1}\\\nonumber 
=
\psi^{n}
+ {\Delta L^{2}}\omega^{n}
-\frac{\Delta L^{2}}{r^2}\psi^{n}
\\
+\Delta L^4\left(\frac{\partial^2 }{\partial r^2}
+\frac{1}{r}\frac{\partial }{\partial r} 
\right)
\left(\frac{\partial^2}{\partial z^2} 
\right)\psi^{n},
\label{psi-equation2}\\\nonumber
 \left[ 1 - \Delta L^{2}\left(\frac{\partial^2 }{\partial r^2}
+\frac{1}{r}\frac{\partial }{\partial r} 
+ \frac{1}{\mu}{\frac{\partial \psi^n }{\partial z} }\frac{\partial }{\partial r}
\right)\right]
\left[1-\Delta L^{2}\left( \frac{\partial^2 }{\partial z^2}\right.\right.
\\\nonumber
\left.\left.-\frac{1}{\mu}\left(  \frac{\partial \psi^n }{\partial r} + \frac{\psi^n }{r} \right)
\frac{\partial}{\partial z}\right)\right]\omega^{n+1}
=\omega^{n}
-\left(\frac{\Delta L^{2}}{ r^2}
\right)\omega^{n}
\\ \nonumber
- {\Delta L^{2}}K_1 \omega^{n}  
+ {\Delta L^{2}}K_2 \omega_s^{n}		      
\\ \nonumber
+ \Delta L^4 \left(\frac{\partial^2 }{\partial r^2}
+\frac{1}{r}\frac{\partial }{\partial r} 
+ \frac{1}{\mu}{\frac{\partial \psi^n }{\partial z} }\frac{\partial }{\partial r}
 \right)                                                  \\
\left( \frac{\partial^2 }{\partial z^2}
- \frac{1}{\mu}
\left(  \frac{\partial \psi^n }{\partial r} 
+ \frac{\psi^n }{r} \right) 
\frac{\partial}{\partial z}\right)\omega^{n}. 
\label{omega-equation2}
\end{eqnarray}
}

In order to compute the solutions on a two dimensional grid, an initial 
guess on $\psi_{ij}^{1}$ and $\omega_{ij}^{1}$ is made for stream function 
and vorticity respectively, by additionally imposing the desired boundary 
conditions. In each iteration with index $n$, beginning from $n=1$,  
Eq.~(\ref{psi-equation2}) is first solved for the radial 
operator part (i.e., r-part) on $A_{ij}^{n+1}$, which formally represents the 
result of the second factor in the LHS of 
Eq.~(\ref{psi-equation2}) operating on the updated stream function $\psi^{n+1}$,
{\small
\begin{eqnarray}
\left[1 -\Delta L^{2}\frac{\partial^2 }{\partial z^2}\right] 
\psi_{ij}^{n+1}=A_{ij}^{n+1}.
\label{def-A}
\end{eqnarray}
}
The values of the computed radial operator part
$A_{ij}^{n+1}$ allow the determination of $\psi_{ij}^{n+1}$ 
by inverting the axial operator part (i.e., z-part) in Eq.~(\ref{def-A}). 
The advantage of this process is that the above equations
(\ref{psi-equation2}) and (\ref{def-A}) are reduced to tridiagonal
systems which is numerically more efficient~\cite{press1992art}. 

\begin{figure}
    \includegraphics{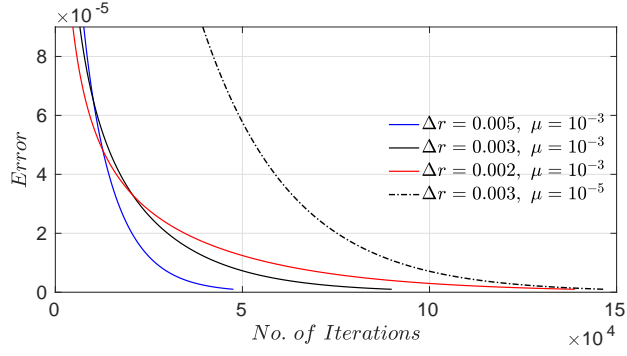}
      \caption{\small Relative error as a function of iterative steps near the 
      tolerance limit of $10^{-6}$ for various grid size $\Delta r[L_r]$ and 
      kinematic viscosity $\mu [U_{0}L_{r}] $.
 \label{fig_error}
 }
\end{figure}

An identical procedure is applied for determining $\omega_{ij}^{n+1}$ 
by defining
{\small
\begin{eqnarray}
\left[1-\Delta L^{2}\left( \frac{\partial^2 }{\partial z^2}
-\frac{1}{\mu}\left(  \frac{\partial \psi^n }{\partial r} 
+\frac{\psi^n }{r} \right)
\frac{\partial}{\partial z}\right)\right]\omega_{ij}^{n+1}
=B_{ij}^{n+1}.
\label{def-B}
\end{eqnarray}
}
Then Eq.~(\ref{omega-equation2}) is first solved for $B_{ij}^{n+1}$, which 
allows Eq.~(\ref{def-B}) to be solved for $\omega_{ij}^{n+1}$. The
updated $\psi_{ij}^{n+1}$ values are used in second half of the iteration
to compute $B_{ij}^{n+1}$ and $\omega_{ij}^{n+1}$, rather than 
the old values $\psi_{ij}^{n}$, which concludes the $n^{\rm th}$ iteration. 
The iterations are made updating the $\omega$ and $\psi$ fields until 
the minimum values of the residues $R_{1} (=\frac{\Delta \psi}{\Delta L^2})$ and 
$R_{2} (=\frac{\Delta \omega}{\Delta L^2})$ below a reasonably small tolerance 
are achieved. The relative 
change in errors define by $Error=max(|\frac{\omega^{n+1}-\omega^{n}}{\omega^{n}}|)$
decreases in each iterative step down to the tolerance limit ($10^{-6}$), which 
ensures the convergence to a steady state solution as shown in Fig~\ref{fig_error}.
The iteration step ${\Delta L}$, grid size $\Delta r$, and kinematic viscosity $\mu$ 
are the main parameters that affect the speed of convergence and numerical stability.
 \section{Characterization of steady equilibrium dust flow structure.}
 \label{characterization}
The above equations are solved subject to proper boundary conditions. 
The boundary for the dust flow in
the present treatment is defined by the effective potential that allows no 
dust flow across the confined domain where the potential $V_b$ for dust jumps
from a small accessible value to a large value at boundaries.
This confinement property($u_{\perp}=0$) at the boundary gives the boundary
conditions for stream function say $\psi_{wall} = 0$. 
The boundary condition for $\omega_{wall}$ depends on the 
nature of $u_{\parallel}$ of the confined dust flows. 
A perfect slip, partial slip, and no slip are
some of the common boundary conditions for $u_{\parallel}$. 
The difficulty with a vorticity stream function
formulation is the lack of natural boundary conditions in term of 
vorticity $~\omega_{wall} $, which however can be derived approximately using 
Thom's formula~\cite{weinan1996vorticity}.

For easy comparison with the analytic solutions presented 
in~\cite{doi:10.1063/1.4887003,PhysRevE.91.063110}, we have used the identical 
conditions as the typical laboratory glow discharge argon plasma 
with micron-sized dust. The plasma parameters are $n \simeq 10^{9}$ cm$^{-3}$,
$T_{e}\simeq 3 eV$, and $T_{i}\simeq 1 eV$ largely at the sheath entrance 
where ions are streaming with a flow velocity $U_{0}$ at the order of the ion 
acoustic velocity $c_{s}=\sqrt{T_{e}/m_{i}}$. 
Further, using the radial width of the confined domain $ L_r$ and steaming ions 
velocity $U_{0}$ as the ideal normalization units    
for the lengths and velocity of the dust flow system, the 
value of ion drag coefficient can be estimated as 
$\xi\sim 10^{-5}~ [U_{0}/L_{r}]$ and the neutral collision frequency
as $\nu\sim 10^{-1}~ [U_{0}/L_{r}]$ 
\cite{PhysRevLett.68.313,PhysRevE.66.046414,PhysRevLett.92.205007}.
For a typical system size, $L_r \sim 10 ~cm $, the range of kinematic 
viscosity $\mu$ can similarly be chosen as $\mu\sim 10^{-6} ~ [U_{0}L_{r}]$ 
which corresponds to the small Reynolds numbers (Re$~\approx 1$) of 
the dust flow in the linear viscous regime~\cite{PhysRevE.91.063110}.

\subsection{Benchmark of the numerical solutions }
\label{analysis_1}
\begin{figure}
    \includegraphics{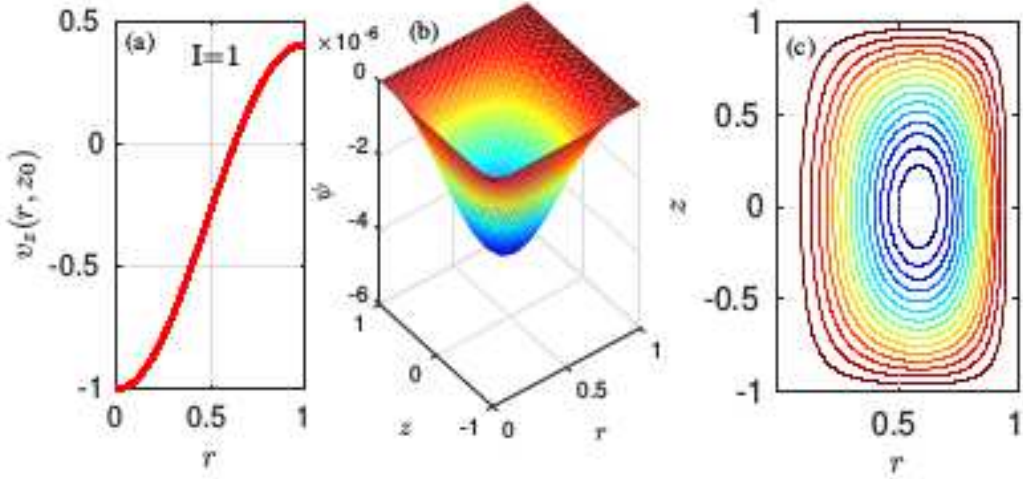}
      \caption{\small 
      (a) Driver ion velocity flow profile $v_z(r,z)$ with axial and radial
      shear at $z=z_0$ cross-section, 
      (b) 2D stream function $\psi(r,z)$ of the dust flow, and
      (c) Corresponding streamlines for the dust fluid flow in $r-z$ 
      crossection for system parameters
           $\mu=~1\times10^{-3} [U_{0}L_{r}] $, 
	   $\xi=~1\times10^{-5} [U_{0}/L_{r}] $ and
           $\nu=~1\times10^{-1} [U_{0}/L_{r}] $.
 \label{fig_2_psi}
 }
\end{figure}

The present model for dust clouds is generic and and applicable to many
driven-dissipative dynamic equilibrium system. It has the freedom of choice for 
external driver, i.e., the background ions flow, and the boundary of 
confining domain. However, for direct comparison with the 
previous analytic solutions~\cite{PhysRevE.91.063110}, now we consider the 
same physical conditions, especially the same external driving field. 
The flow profile of streaming ions is specified as, 
 \begin{eqnarray}
v_{z}(r,z)= A_{m}J_0\left(\alpha_{I} \frac{r}{L_{r}}\right)
\cos{\left(\frac{\pi z}{2 L_{z}}\right)}.
\label{vz_bessel_sm}
\end{eqnarray}
Here $A_{m}$ is the magnitude of external driving field. The
radial modes are determined by the $I_{th}$ root of Bessel function $J_0(\alpha_I)=0$ 
at the external boundary. The axial mode $k_z =(\pi/{2L_z})$ specifies 
a single vortex along axial direction as in experimental
observations~\cite{doi:10.1063/1.4929916}.
The numerical solution for the stream function $\psi(r,z)$ and its corresponding 
streamlines ( i.e., the contours of the product $ r\psi $) in the linear 
viscous regime are shown in Fig.~\ref{fig_2_psi}(b)-(c) respectively for
$\mu = 10^{-3} [U_{0}L_{r}] $, $ \xi = 10^{-5} [U_{0}/L_{r}]$ 
and $\nu=  10^{-1} [U_{0}/L_{r}] $. 
This streamline pattern shows similar characteristic features of 
low-Re dust flow as in~\cite{PhysRevE.91.063110}, where the flow is 
anti-clockwise circulation, axial symmetric and aligned 
to the confining boundary, ensuring vorticity transport purely due to 
diffusion orthogonal to the streamlines.

\begin{figure}
      \includegraphics{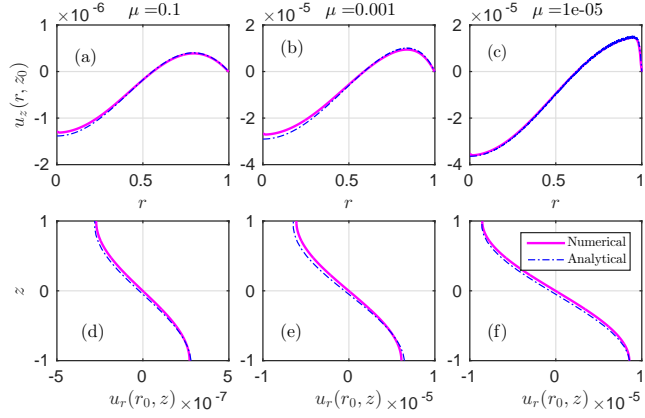}
      \caption{\small Comparison between analytical and numerical 
      solutions of $ u_{r}(r_0,z)~and~u_{z}(r,z_0)$ for fixed system 
      parameters, 
        $\xi=~ 1\times10^{-5}[U_{0}/L_{r}] $,
        $\nu=~ 1\times10^{-1}[U_{0}/L_{r}] $,
        and varying $\mu=~ 10^{-1} ~to ~ 10^{-5}[U_{0}L_{r}]$ 
        in the linear regimes Re$\ll1$.
 \label{fig_3_cmp}
 }
\end{figure}

A more quantitative comparison is possible between the flow velocity 
component profiles obtained numerically and those from 
analytic solutions in the linear limit~\cite{PhysRevE.91.063110}.
The profiles of the dust flow velocity components, i.e., $ u_{z}(r,z_0)$ and
$ u_{r}(r_0,z)$ passing through the center of the circulation $(r_0,z_0)$ are 
compared for the kinematic viscosity varying 
from $\mu =10^{-1} ~to ~\mu =10^{-5}[U_{0}L_{r}]$ in the 
linear regime (Re$\ll1$) (Fig.~\ref{fig_3_cmp}).
In the highly viscous regimes, the dust fluid flows with a velocity in the 
range of $0.1-1 ~mm/sec$ in the confined domain. 
The analytical and numerical results are in good agreement.
The above comparisons show the validity of the
numerical method and motivates for further analysis of the driven dust flow
characteristics at higher Reynolds number nonlinear regimes. 
Further, the variations in velocity profiles for $ u_{r}(r_0,z)$ and $ u_{z}(r,z_0)$ 
near the boundary indicates the impact of boundary conditions 
and the formation of boundary layer in particular. Boundary layers are formed due to 
the effective viscous stress on flow due to boundaries. 
The thickness of the boundary layer decreases with kinematic viscosity as 
shown in Fig.~\ref{fig_3_cmp}(a)-(c). 
Thus at higher Reynolds number, the thickness of boundary layer is negligibly 
small as $\Delta r_{b} \simeq \mu^{1/3}$~\cite{PhysRevE.91.063110}, 
and reduces to a very thin layer giving sharp deviation in velocity profile 
as shown in Fig.~\ref{fig_3_cmp}(f). There is no boundary layer formation 
for the perfect slip boundary conditions as shown in Fig.~\ref{fig_3_cmp}(d)-(f).

\subsection{Emerging nonlinear characteristics in the steady equilibrium dust 
flow structures}
\label{analysis_2}
When the contribution of nonlinear advection transport ${\bf(u\cdot\nabla)u}$ is 
included in the higher Reynolds number regimes, the momentum of dust flow 
increases enormously and it becomes important to maintain a low driver 
ion velocity so that Mach number of the 
dust $ M_d = u_{d}/c_{sd} \ll 1 $ for incompressible flow 
in practice. Thus, for the study of nonlinear effects at higher Re regimes, 
the driver ions stream with a shear flow velocity $U_0$ equivalent to a 
fraction of the ion acoustic speed $c_s$. The order of kinematic 
viscosity corresponding to small Reynolds numbers (Re$\simeq 1) $ 
of dust flow becomes  $\mu\sim 10^{-4} ~ [U_{0}L_{r}]$ which 
is consistent with the linear viscous regime. The corresponding ion drag coefficient 
can be estimated as $\xi\sim 10^{-4}~ [U_{0}/L_{r}]$, and 
neutral collision frequency as $\nu\sim 10^{-3}~ [U_{0}/L_{r}]$ for 
further analysis~\cite{PhysRevLett.68.313,PhysRevE.66.046414,PhysRevLett.92.205007}.

\begin{figure}
    \includegraphics[width=10cm]{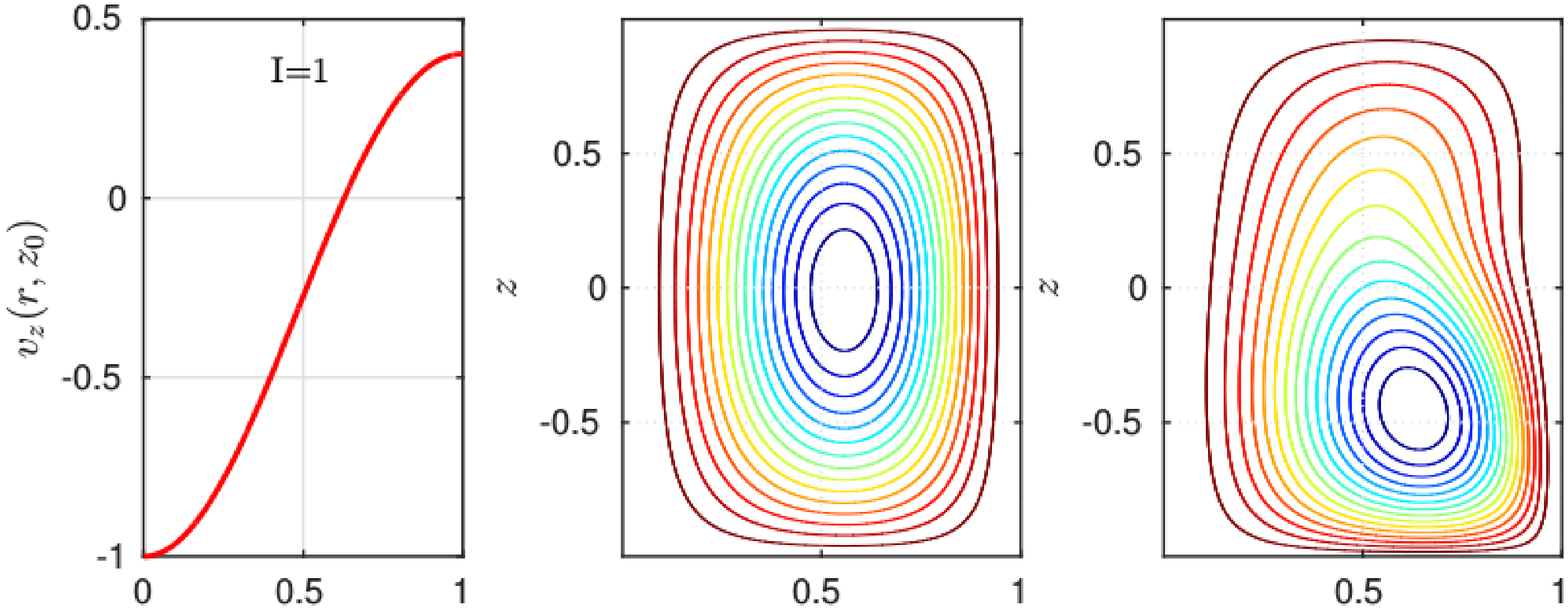}\\
    \includegraphics[width=10cm]{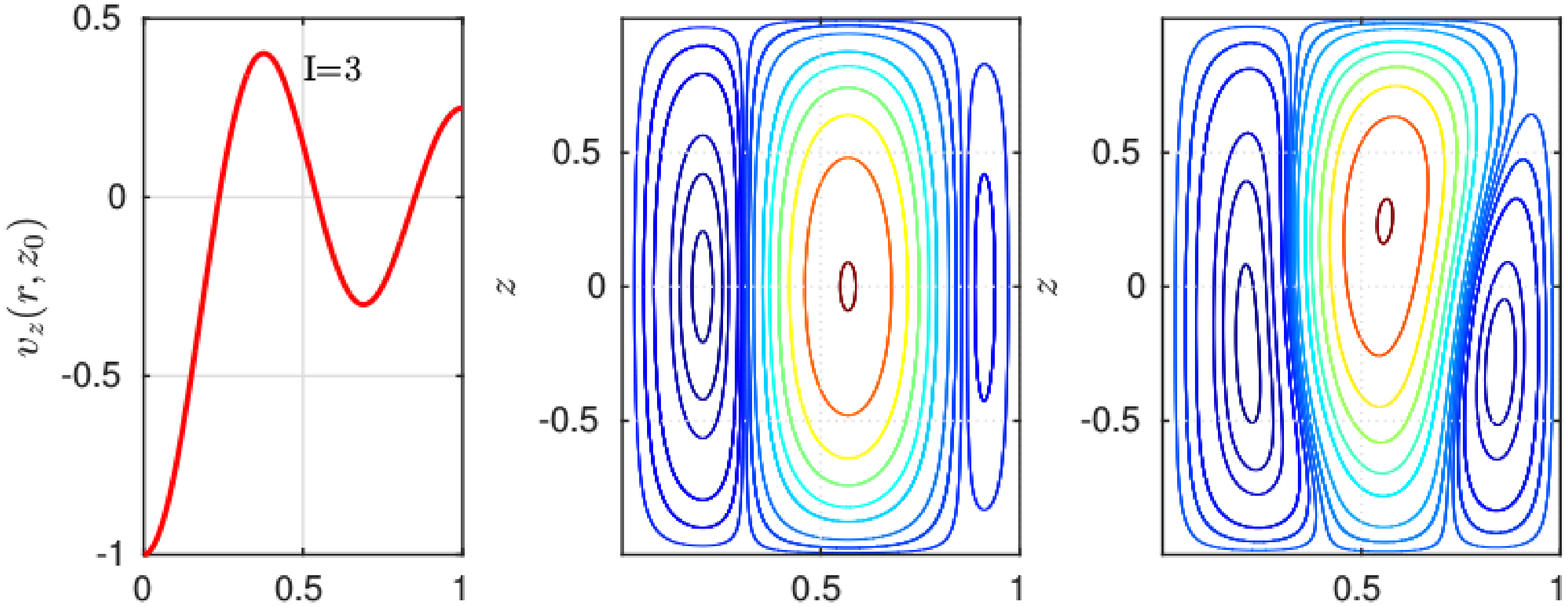}
 \caption{\small
 ($1^{st} ~coulum$) Driver ion's velocity profile at $z=z_0$ for radial 
  sheared mode numbers $I=1~and~3$. Corresponding streamlines pattern for the dust 
 fluid flows in $r-z$ crossection for ($2^{nd}~coulum$) 
 $\mu = 10^{-3} [U_{0}L_{r}] $, and
 ($3^{rd}~coulum$) $\mu = 3\times10^{-5} [U_{0}L_{r}] $ respectively.
 \label{fig_diff_modes}
}
\end{figure}

Now the above flow analysis of the bounded driven dust flow dynamics is extended 
to high Re regimes. 
The streamlines of 2D steady dust flow in the linear and nonlinear regimes are presented 
in Fig.~\ref{fig_diff_modes} for the radial mode numbers $I=1$ and $3$ of the 
same driver flow.
The flow structures are analyzed for two important cases of 
kinematic viscosity, i.e., $\mu = 10^{-3} [U_{0}L_{r}] $ for the linear regime  
and $\mu = 10^{-5} [U_{0}L_{r}] $ for the nonlinear regime which corresponds  
to flow regimes Re$~\approx 0.001$ and Re$~\approx 50.0$ respectively. 
In these regimes, the dust fluid flows within the velocity range 
of $1-50 ~mm/sec$, while the dust acoustic speed $c_{ds}$ is found to be $12 ~cm/sec$ 
in the domain. 
As observed in the earlier analysis~\cite{PhysRevE.91.063110,
doi:10.1063/1.4887003}, the flow structure in the linear regime is characterized by 
a system of symmetric and elongated circulations (or streamlines that are aligned 
with the confining boundaries) as shown in Fig.~\ref{fig_diff_modes} ($2^{nd}~coulum$).
The alternating positive and negative shear in the external driving field in case 
of $I=~3$ leads to the formation of counter-rotation in addition to scales introduced by 
the confining boundary and stationary background fluid~\cite{PhysRevE.91.063110}. 
However in the nonlinear regime, as shown in Fig.~\ref{fig_diff_modes}($3^{rd}~coulum$), 
the flow structure changes into an asymmetric and elongated circulation 
with a finite displacement of the 
center of circulation ${\bf( u(r_0,z_0)=0)}$. 
In the case of multiple vortices (in the $I=~3$ case), the circulation center 
shifts along with the driver. 
In many dusty plasma experiments~\cite{1367-2630-5-1-366,doi:10.1063/1.4929916},
similar multiple dust vortices are observed to correlate with scales of 
the ion dragging interaction in background.
Further, a remarkable property of the present analysis is that the geometry and 
the dimension of steady driven dust vortex structure in the linear regime are mainly 
asserted by the confining boundary whereas in the nonlinear regime, the structure is mostly
controlled by the dynamical regime rather than the boundary. 
Such vortex dynamics and structural changes are observed in recent dusty plasma 
experiments. For example, S. Mitik{\it~et.~al.}~\cite{PhysRevLett.101.235001} 
observed dust vortex driven by neutral gas convection, and structural variation 
with change in pressure in the applied field. In another experiment, 
T. Hall{\it~et.~al.}~\cite{Hall2016ASO} observed the asymmetric response 
of dust clouds to a change in the relative strengths of the electrostatic 
and ion drag forces at varying pressure.
\subsection{Flow structure dependence on the dynamical regime 
  and domain aspect ratio.}
  \label{analysis_3}
The influence of various system parameters or dynamical regimes 
on vortex structure is further analyzed explicitly for the $I=1$ case
(Fig.~\ref{fig_diff_modes},~$1^{st}~row$).  
For the driven-dissipative system, the steady state momentum equation 
can be written as
\begin{eqnarray} \nonumber
({\bf u} \cdot \nabla) {\bf u}  
= -\frac{\nabla{P}}{\rho} -{\nabla{V}} + \mu \nabla^2 {\bf u} 
-\xi({\bf u}-{\bf v_s})-\nu({\bf u}-{\bf w_n}).    
\label{ns-equation1}
\end{eqnarray} 
Applying the condition for the center of circulation, i.e., ${\bf u(r_0,z_0)=0}$, 
the equation at ${(r_0,z_0)}$ reduces to  
\begin{eqnarray} 
\frac{\nabla{P}}{\rho}+{\nabla{V}}- \mu \nabla^2 {\bf u}
=\xi{\bf v_s}+\nu{\bf w_n}.    
\label{ns-equation11}
\end{eqnarray}
Here, $P$ is the dynamic pressure of the incompressible flow which becomes a 
function of velocity ${\bf u}$, which again depends on $\mu$. 
V is the effective confining potential and the diffusion transport 
$\mu \nabla^2 {\bf u}$ at the center point $(r_0,z_0)$ is negligibly small. 
\begin{figure}
    \includegraphics{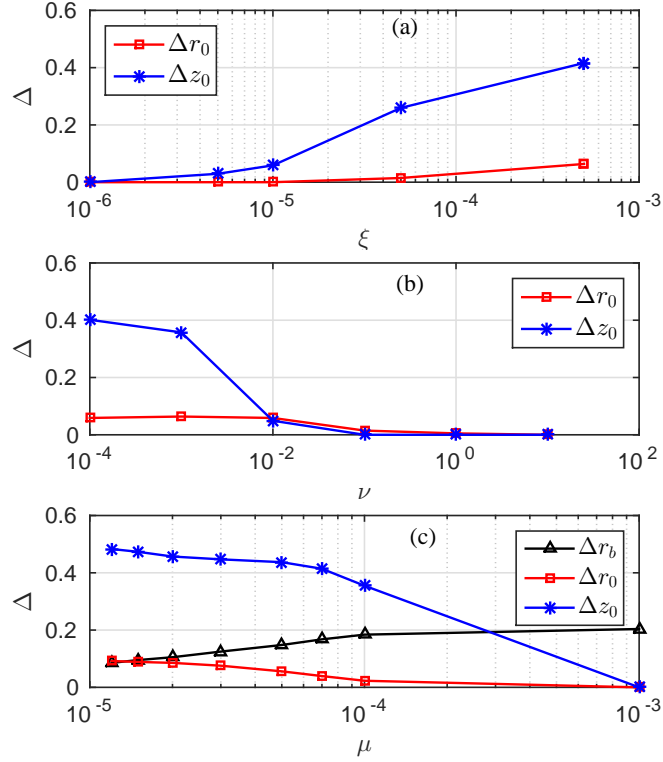}
 \caption{\small 
   Shift of circulation center ($r_0,z_0$) as function of (a) ion dragging 
   co-efficient $\xi$, (b) neutral collision frequency $\nu$, and
   (c) kinematic viscosity $\mu$. The boundary layer 
   thickness $\Delta r_b$ as function of $\mu$ is also plotted in (c).
 \label{fig_shift}
}
\end{figure}
Now, Fig.~(\ref{fig_shift})(a) and (b) shows the radial and axial shifts of 
the center point $(r_0,z_0)$ with respect to $\xi$ and $\nu$ 
respectively. 
The radial shift $\Delta r_0$ is small in each case, while the axial 
shift $\Delta z_0$ with $\xi,~\nu$ is high in the higher Reynolds regime. 
However in the case of $\mu$ variation shown in Fig.~(\ref{fig_shift})(c), 
the shift $\Delta z_0$ increases slowly in the higher Re regime. 
Further, the radial shift $\Delta r_0$ and the change in boundary layer thickness 
$\Delta r_b$ due to varying kinematic viscosity $\mu$ correlate with each other. 
The different variations in radial and axial shifts of the center of circulation 
$(r_0,z_0)$ with respect to system parameters suggest that the steady flow 
structure also depends on the aspect ratio ($L_z/L_r$) of the confined domain.
This may be because in the nonlinear regime, $({\bf u} \cdot \nabla) {\bf u}$ introduces 
new scale $L_{\|}\sim(u/u')$ in addition to the diffusion scale ($L_{\perp}$).
The analysis of nonlinear flow characteristics in a domain of unity aspect ratio  
($L_z/L_r =1$)\cite{PhysRevE.95.033204,doi:10.1063/1.5020416} shows that the shifting in 
both the $\Delta z_0$ and $\Delta r_0$ are relatively small and similar. 
\section{Summary and conclusions}
\label{conclusion}
A 2D hydrodynamical model is developed for the dynamics of a 
dust fluid embedded in a combination system of an unbounded sheared 
ion flow and a background stationary neutral fluid. 
The validity of the present numerical solution is verified 
by comparing the flow profiles produced by the previous analytical solution in 
the linear limit\cite{PhysRevE.91.063110}. 
Improving from the earlier 
analysis\cite{doi:10.1063/1.4887003,PhysRevE.91.063110}, this model has 
the freedom of choosing any form of the driving flows, confining boundaries, 
and arbitrary Reynolds number regimes.
The vortex structure of the steady dust fluid in the poloidal cross-section is 
analyzed within the transition range of Reynolds number $0.001< Re \le 50 $ 
where the dust flow with speed ranges from $0.1~mm~to~3~cm$ while the dust acoustic
speed is about $12~cm/sec$. It has been observed that the geometry and dimension of 
the steady dust vortex in the linear regime (Re$\le1$) is mainly determined 
by confining boundaries, whereas it is governed by the dynamics (or dynamical regime) 
rather than the boundaries in the nonlinear regime (Re$\gg1$). 
Thus the steady dust flow structure in the linear regime is characterized by 
symmetric and elongated vortices with various scales mainly introduced by the 
driving fields and boundaries. However at Re$\gg 1$, the steady dust flow 
structure is no longer symmetric. Instead, a new asymmetric and elongated structure 
appears with scales mainly dependent on the driven-dissipation parameters such as 
$\xi$, $\nu$, and $\mu$ in addition to the aspect ratio of the confined domain. 

It is concluded that steady driven and bounded dust flow in an 
unbounded streaming plasma can possess various flow structures in linear and 
nonlinear dynamic regimes depending on scales introduced by non-uniformity of the 
unbounded external driving field and the system parameters including aspect ratio and 
those of confining boundaries. The numerical model is generic and 
applicable for other driven-dissipative systems such as the microscopic biological 
system and the gigantic Jovian vortices. 
Further analysis on the nonlinear flow characteristics and its stability in various 
aspect ratios and arbitrary Re regimes will be reported in future publications. 
\section*{Acknowledgement}
Author L.~Modhuchandra acknowledges Dr. Devendra Sharma and late Prof. P. K. Kaw, 
Institute for Plasma Research, India, for the invaluable support and encouragement all the time. The research was supported by the State Administration of Foreign Experts Affairs - Foreign Talented Youth Introduction Plan Grant No. WQ2017ZGKX065, the National Magnetic Confinement Fusion Program of China Grant Nos. 2014GB124002 and 2015GB101004.
Author P.~Zhu also acknowledges the supports from U.S. Department of Energy Grant
Nos. DE-FG02-86ER53218 and DE-FC02-08ER54975.  
This work used resources of Supercomputing Center of University of Science and Technology of China. 
%

\end{document}